\documentclass[5p,authoryear]{elsarticle}
\usepackage{fancyvrb}
\usepackage[utf8x]{inputenc}
\usepackage{natbib}
\usepackage{amssymb}
\usepackage{amsmath}
\usepackage{url}
\usepackage{caption}
\usepackage{subcaption}
\usepackage{graphicx}
\usepackage{color}
\usepackage{microtype}
\usepackage{booktabs}
\usepackage[hidelinks]{hyperref}
\usepackage{enumerate}

\journal{Astronomy and Computing}

\begin{document}

\newcommand{\code}[1]{\texttt{#1}}
\newcommand{\degrees}[1]{$#1^\circ$}

\newcommand{\githash}{6b20b9d}\newcommand{\gitdate}{2015-08-25}

\begin{frontmatter}

\title{\texttt{Chimenea} and other tools: \\ Automated imaging of multi-epoch radio-synthesis data with CASA \tnoteref{git}}
\tnotetext[git]{Generated from git source \texttt{\githash} dated \gitdate.
Source available at \href{https://github.com/timstaley/automated-radio-imaging-paper}{github.com/timstaley/automated-radio-imaging-paper}
}

\author[oxford]{Tim D. Staley}
\ead{tim.staley@physics.ox.ac.uk}
\author[oxford]{Gemma E. Anderson}

\address[oxford]{Department of Physics, University of Oxford, Denys Wilkinson Building, Keble Road, Oxford, OX1 3RH, UK}

\begin{abstract}
In preparing the way for the Square Kilometre Array and its pathfinders, there is a pressing need to begin probing the transient sky in a fully robotic fashion using the current generation of radio telescopes. Effective exploitation of such surveys requires a largely automated data-reduction process.
This paper introduces an end-to-end automated reduction pipeline, \texttt{AMIsurvey}, used for calibrating and imaging data from the Arcminute Microkelvin Imager Large Array. 
\texttt{AMIsurvey} makes use of several component libraries which have been packaged separately for open-source release. 
The most scientifically significant of these is \texttt{chimenea}, which implements a telescope-agnostic algorithm for automated imaging of pre-calibrated multi-epoch radio-synthesis data, of the sort typically acquired for transient surveys or follow-up.
The algorithm aims to improve upon standard imaging pipelines by utilizing iterative RMS-estimation and automated source-detection to avoid so called `Clean-bias', and makes use of CASA subroutines for the underlying image-synthesis operations.
At a lower level, \texttt{AMIsurvey} relies upon two libraries, \texttt{drive-ami} and \texttt{drive-casa}, built to allow use of mature radio-astronomy software packages from within Python scripts. 
While targeted at automated imaging, the \texttt{drive-casa} interface can also be used to automate interaction with \textit{any} of the CASA subroutines from a generic Python process.
Additionally, these packages may be of wider technical interest beyond radio-astronomy, since they demonstrate use of the Python library \texttt{pexpect} to emulate terminal interaction with an external process. 
This approach allows for rapid development of a Python interface to any legacy or externally-maintained pipeline which accepts command-line input, without requiring alterations to the original code.

\end{abstract}


\end{frontmatter}

\section{Introduction}
\label{sec:intro}
The science of transient astronomical events is a burgeoning sub-field of astronomy. 
For the past decade, optical and gamma-ray transient surveys have led the way, with radio-band observations being largely restricted to follow-up of manually selected targets discovered at other wavelengths.
However, a new generation of radio facilities composed primarily of Square Kilometre Array (SKA) pathfinders such as 
ASKAP \citep[the Australian Square Kilometre Array Pathfinder;][]{Murphy2013},
MeerKAT \citep[Karoo Array Telescope;][]{Booth2012}, 
LOFAR \citep[the Low Frequency Array;][]{vanHaarlem2013}, 
MWA \citep[Murchison Widefield Array;][]{Tingay2013,Bell2014}, 
and the Jansky Very Large Array (JVLA), are now at various stages of commissioning and planning with some (LOFAR, JVLA, MWA) already producing new surveys and the rest scheduled for commissioning within the next few years \citep[see e.g.][for an overview]{Norris2013}.
These new observatories, and in time the SKA, will provide wider fields of view, 
greater sensitivity and potentially much faster response to targets of opportunity than was previously possible. 
This makes them attractive instruments for performing transient science, both for routine multi-wavelength follow-up and a new generation of dedicated surveys in the radio-band \citep{Stappers2013}. 
Estimates of transient-discovery rates from SKA surveys currently vary quite widely
\citep[see e.g.][for orphan gamma-ray burst rate estimates that differ by an order of magnitude, primarily due to different choices of signal-to-noise thresholds]{Burlon2015,Metzger2015}, but it is safe to assume that the total number of transient alerts from across the spectrum will only increase over the coming years, primarily due to optical survey projects such as the Large Synoptic Survey Telescope \citep{Ridgway2014}. 
Exploiting the potential of the the SKA and its pathfinders for transient science requires that we follow-up these transient alerts in much greater numbers compared to current practice. As such there is a pressing need to begin probing the transient sky in a fully-robotic fashion using the current generation of radio telescopes, in order to develop the techniques and infrastructure that will be required.

Since 2012 we have been using the Arcminute Microkelvin Imager Large Array \citep[AMI-LA,][]{Zwart2008} to conduct a programme of automated radio-followup of transients detected at other wavelengths, now known as the AMI-LA Rapid Response Mode (ALARRM).
ALARRM has produced some of the earliest 
gamma-ray burst (GRB) follow-up in the radio \citep{Staley2013,Anderson2014},
and unprecedented early-time coverage of a radio-flare from an M-dwarf
star accompanying a hard X-ray outburst \citep{Fender2015}.

Existing facilities such as AMI-LA typically rely upon mature data-reduction pipelines created with an interactive, user-intensive model in mind, making them ill-suited for integration into fully-automated analyses of the sort needed for a quick evaluation of transient candidates. 
However, these legacy pipelines are the result of many developer-years of effort in terms of algorithm development, software engineering, and testing.
As such they represent an irreplaceable part of the data-reduction process for existing observatories. 

In the course of the ALARRM programme we have developed modular software, written in Python, to automate the data-reduction process. 
We have recently open-sourced this software as four separate packages.\footnote{%
The packages described in this paper are available from\\
\url{https://github.com/timstaley/drive-ami},\\
\url{https://github.com/timstaley/drive-casa},\\
\url{https://github.com/timstaley/chimenea}, and\\
\url{https://github.com/timstaley/amisurvey}.
}
The first two, \texttt{drive-ami} and \texttt{drive-casa} \citep{Staley2015_amisurvey, Staley2015_drivecasa}, are libraries developed to provide interfaces to mature radio-astronomy software packages, to enable their use from within Python scripts.
A third library, \texttt{chimenea} \citep{Staley2015_chimenea}, implements an heuristic algorithm for automated imaging of pre-calibrated multi-epoch radio-synthesis data, using \texttt{drive-casa} to build upon subroutines from CASA \citep[the Common Astronomy Software Applications package,][]{McMullin2007, Casa2011}.
The final package, \texttt{AMIsurvey} \citep{Staley2015_amisurvey}, ties together the previous codes to create a telescope-specific end-to-end pipeline, which ingests raw AMI-LA datafiles and produces calibrated and cleaned multi-epoch images ready for source extraction and transient identification.

This paper is structured as follows: 
In Section~\ref{sec:drive-ami} we introduce AMI-LA, and describe a new fully scriptable Python interface to its associated calibration pipeline.
Section~\ref{sec:drive-casa} discusses our choice to build upon the CASA routines to implement the imaging stage, and explains the rationale and design of our external
scripting interface, \texttt{drive-casa}.
Section~\ref{sec:chimenea} describes the automated imaging algorithm implemented in the \texttt{chimenea} package, and Section~\ref{sec:amisurvey} gives an overview of the end-to-end ALARRM data-reduction pipeline encoded by \texttt{AMIsurvey}.
Section~\ref{sec:results} presents a basic performance analysis demonstrating that the pipeline behaves as expected.
In the discussion section, we describe known limitations of our \texttt{chimenea} algorithm, and cover some possible extensions and alternatives (\S\ref{sec:chimenea-discussion}). 
We then describe and compare different methods for solving a problem relevant in the wider context of software in academia --- that of interfacing with legacy software --- and explain the choice of method for this work (\S\ref{sec:software}). 
We finish the discussion with some general recommendations for writing re-usable Python codes (\S\ref{sec:code-reuse}), and conclude in Section~\ref{sec:conclusion}.

\section{Automated calibration of AMI-LA data with \texttt{drive-ami}}
\label{sec:drive-ami}
The Arcminute Microkelvin Imager Large Array (AMI-LA) is a synthesis telescope composed
of eight equatorially mounted 12.8 m dishes sited at the Mullard
Radio Astronomy Observatory (MRAO) at Lord’s Bridge, Cambridge. The
telescope observes in the band 12.5–18.2 GHz with eight 0.72 GHz
bandwidth channels. 

AMI-LA observations are recorded in a custom data format to be processed by a specialized pipeline, \texttt{AMI-REDUCE}. 
This applies path delay corrections, automatic flags for interference, pointing errors, shadowing and hardware faults, applies phase and amplitude calibrations, Fourier transforms the data into the frequency domain, and writes out the resulting data in \textit{uv}-FITS format \citep{Davies2009}. 
Typically data reduction is either carried out interactively, allowing the user to view plots of the data in order to guide the reduction process, or by calling one of a selection of standard reduction scripts (which are simple listings of \texttt{AMI-REDUCE} commands, without any flow control such as `if' statements or `for' loops.). 

Application of one of the standard \texttt{AMI-REDUCE} listings was adopted as a calibration step for observations made as part of the ALARRM programme. 
Early versions of the ALARRM pipeline ran \texttt{AMI-REDUCE} from a Python script via a simple subprocess call (cf \S\ref{sec:subprocess}), but this proved problematic. 
First, when inspecting the logs, a successful reduction with \texttt{AMI-REDUCE} was indistinguishable from an aborted reduction due to a problematic dataset. 
Second, important metadata such as the rain modulation (which gives some idea of the data-quality) and observation time is not stored in the header of the \textit{uv}-FITS output files, and so checking these values still required manually loading the dataset in question into the \texttt{AMI-REDUCE} pipeline. 

Clearly, a more powerful programmatic interface to \texttt{AMI-REDUCE} was required. 
Accordingly, we we implemented an interface that provides a set of Python routines for controlling \texttt{AMI-REDUCE}, under the moniker \texttt{drive-ami}. 
To do so we make use of 
\texttt{pexpect}\footnote{%
\url{http://pexpect.readthedocs.org}}, 
a pure-Python library for emulating terminal-interaction, as discussed in subsections~\ref{sec:emulation} and~\ref{sec:why-emulation}.  

In addition to allowing step-by-step calling of the \texttt{AMI-REDUCE} functionality from a Python script, \texttt{drive-ami} provides a number of enhancements to the data-reduction process.
Data-provenance and reproducibility are improved; all the \texttt{AMI-REDUCE} commands are recorded in a log-file along with the output \textit{uv}-FITS data. 
In a separate file, all textual output displayed during the reduction process is logged alongside the input commands, so that the end-user may read back over the reduction dialogue as if scrolling back through a terminal history arising from using \texttt{AMI-REDUCE} interactively. 
A variety of metadata such as observation times, calibrator source, rain modulation and percent of data flagged due to radio-frequency interference are also parsed from the output and stored in JavaScript Object Notation (JSON) format, which has the benefit of being both human-readable and trivially parsed in most high-level programming languages. 

As an added convenience, \texttt{drive-ami} can also be used to help configure reduction of multiple-epoch data for a source observed with AMI-LA. 
While observations are usually easy to sort by their filename-prefix (which refers to the target name), occasionally a target is renamed after the initial epoch of observation (e.g. to reflect an assigned GRB ID), or a prefix may be incremented if multiple observations are performed in a single day. 
\texttt{drive-ami} can utilize \texttt{AMI-REDUCE} to extract the pointing information from the raw data-files and then group observations according to their pointing, given a tolerance limit on the maximum angular separation between pointing-centres. 
We make use of the \texttt{astropy} co-ordinate routines for calculating the angular separation \citep{Astropy2013}.
The JSON-encoded metadata accompanying the resulting \textit{uv}-FITS data then contains a `group' tag, which can be used to identify all observations of a given target for processing as a group (cf Section~\ref{sec:chimenea}).

\begin{figure*}[p!]
\begin{center}  
  \includegraphics[width=.96\textwidth]{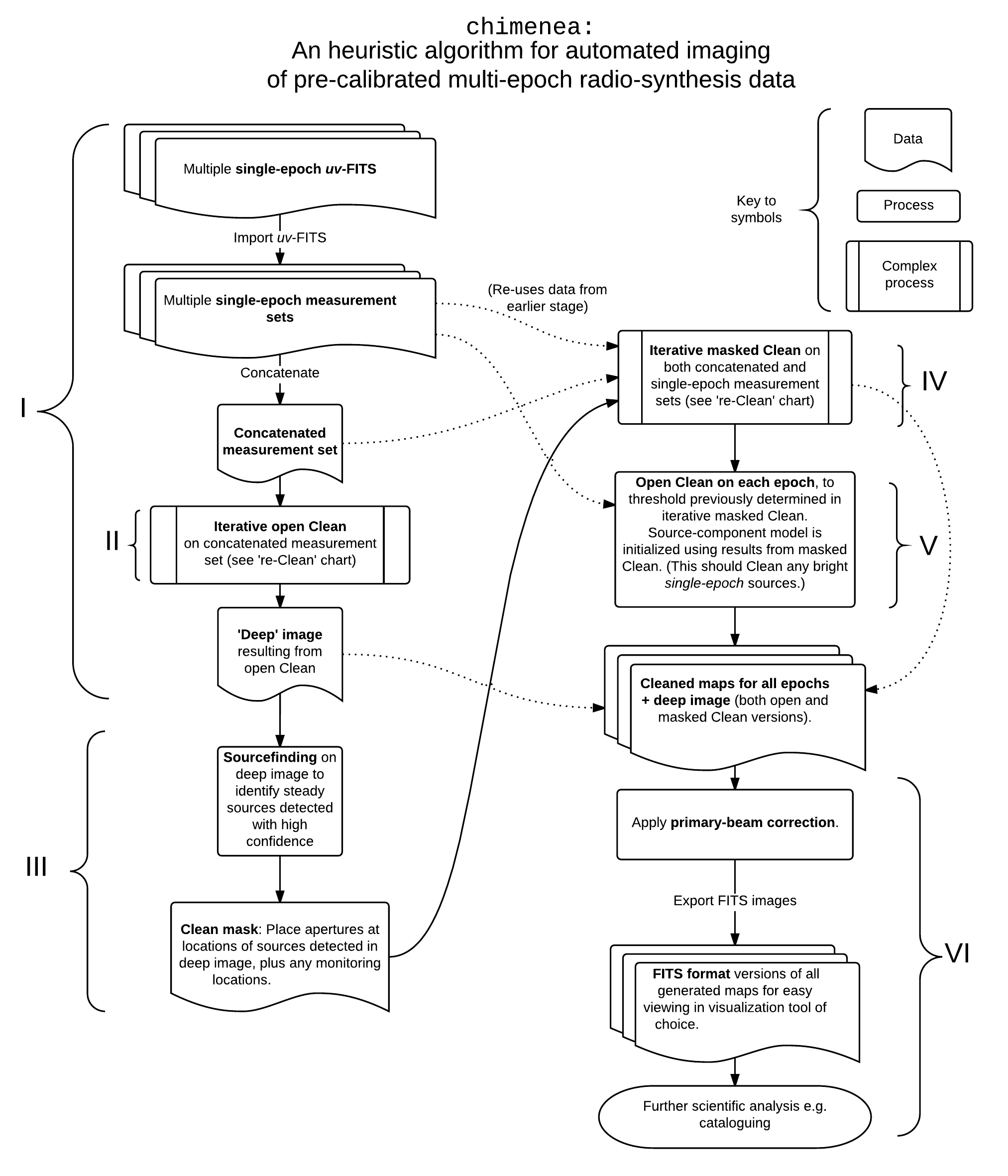}
  \caption[Overview of \texttt{chimenea} logic]{%
An overview of the algorithm implemented in the \texttt{chimenea} pipeline.
Dotted lines denote re-use of data-products from an earlier stage. 
Groupings labelled in Roman numerals refer to descriptions in text.
The terms `open' and `masked' Clean refer to applying the Clean
algorithm \citep{Schwab1984} either to a box-region covering the central quarter of a map (i.e. an open or unconstrained clean), or to a masked map, i.e. one where the Clean algorithm is constrained to placing model components in the vicinity of known sources.
The `re-Clean' subroutine is depicted in Figure~\ref{fig:reclean}.
\label{fig:chimenea}
} 
\end{center} 
\end{figure*}

\section{External scripting of CASA subroutines with \texttt{drive-casa}}
\label{sec:drive-casa}
Traditionally, the \textit{uv}-FITS data products output from the \texttt{AMI-REDUCE} reduction stage are reduced using the \texttt{AIPS} package \citep{Fomalont1981}, but for the ALARRM programme we chose to make use of CASA \citep{McMullin2007}. 
We made this switch for a few reasons.
We judge CASA to be the best supported general-purpose package for radio-astronomy data-reduction, 
with a simple installation process, 
frequent updates\footnote{\url{http://casa.nrao.edu/previous_casa.shtml}}, 
a comprehensive accompanying 
manual\footnote{\url{https://casa.nrao.edu/ref_cookbook.shtml}},
and a helpdesk service. 
CASA is also widely applicable, recognising data from many of the current generation of telescopes; support is also available for newly commissioned facilities such as 
LOFAR\footnote{%
See also the LOFAR imaging cookbook:\\ 
\url{http://www.astron.nl/radio-observatory/lofar/lofar-imaging-cookbook}
}
and ALMA. 
Lastly, the Python interpreter underlying the default CASA interface, \texttt{casapy}, provides a convenient scripting interface for many basic tasks, albeit with some restrictions, as described below.

It is trivial to run simple Python scripts within the \texttt{casapy} environment, or even to launch \texttt{casapy} into a Python script directly from the command line. 
However, problems arise if you wish to make use of functionality from both \texttt{casapy} and non-standard Python packages (i.e., anything except the standard library) from within the same script.
This is because \texttt{casapy} bundles its own Python interpreter, customized to display a logging window and provide access to plotting tools. 
As a result, Python packages installed in the normal manner are unavailable from within the \texttt{casapy} environment. 
This is unfortunate, as it precludes easy usage of CASA routines from within a larger pipeline, even though much of the underlying interface is already written in Python.

With regards to data-provenance and reproducibility, \texttt{casapy} provides extensive logfiles which provide a record of the commands and resulting information from a data-reduction run. 
However, a minor flaw is that once instantiated, \texttt{casapy} provides no method (or at least, no documented method) for redirecting log output, which can make it hard to locate the log section relevant to a given observation when reduced as part of a batch. 
Logging of input and output is also collapsed into a single stream, which makes it tricky to extract the commands required to reproduce a reduction process.

Programmatic error-handling is also lacking from the \texttt{casapy} environment. 
The standard \texttt{casapy} routines do not provide a return value, or throw exceptions on encountering error conditions. 
Instead, all processing information and errors are output as log messages, which makes it hard to respond in an automated fashion.

Our package for automating interaction with \texttt{casapy}, `\texttt{drive-casa}', addresses these issues.
Similarly to \texttt{drive-ami}, we use \texttt{pexpect} to emulate terminal interaction with the \texttt{casapy} process, as detailed in Section~\ref{sec:emulation}. 

The \texttt{drive-casa} interface can be used to execute strings or scripts containing \texttt{casapy} commands directly, so all \texttt{casapy} subroutines are accessible via the \texttt{drive-casa} interface. 
Alternatively, the package includes a set of convenience routines which try to adhere to a consistent style and make it easy to chain together successive CASA reduction commands to generate a \texttt{casapy} script programmatically.
For example, the following code fragment generates a script to invoke the CASA routine \mbox{\textit{importUVFITS}}, then perform a zero-iteration Clean operation on the resulting CASA \textit{MeasurementSet}
to produce a dirty map:

\begin{Verbatim}[samepage=true]
script = []
ms = drivecasa.commands.import_uvfits(
                script, 
                uvfits_path)
dirty_maps = drivecasa.commands.clean(
                script, 
                ms, 
                niter=0, 
                threshold_in_jy=1,
                other_clean_args=clean_args)
                
\end{Verbatim}

We note that the recently released package \texttt{casa-python}\footnote{%
\url{https://github.com/radio-astro-tools/casa-python}
}
should make it easier to install external Python packages into the \texttt{casapy} environment, which goes some way to addressing the issue of package interoperability whilst retaining access to the standard \texttt{casapy} logging interface and plotting tools. 
This may provide a preferable solution for those who wish to make small alterations to their \texttt{casapy} workflow to include external functionality, but still effectively treats \texttt{casapy} as the top-level pipeline framework, rather than abstracting it to a more standard Python package, as \texttt{drive-casa} does.

Since \texttt{casapy} is largely Python code internally, and is still undergoing active development, we hope that eventually a native Python interface to CASA \textit{as a library} (not an executable) will be made available, thus negating the need for the terminal-emulation layer of \texttt{drive-casa}. 
In the meantime, \texttt{drive-casa} provides a reliable method for incorporating CASA functionality into a fully automated Python reduction pipeline, as demonstrated in Section~\ref{sec:chimenea}.

\section{Automated imaging of multi-epoch radio-synthesis data with \texttt{chimenea}}
\label{sec:chimenea}
Prompt analysis of results from the ALARRM programme requires frequent re-reduction of target-datasets as additional observations become available.
In order to minimise the human effort associated with the imaging stage of the data-reduction, we created a tool for automated imaging of pre-calibrated multi-epoch radio-synthesis data, `\texttt{chimenea}', which is a small pipeline built atop the CASA subroutines (using \texttt{drive-casa} to provide an interface layer). 

The \texttt{chimenea} pipeline is designed to automatically choose a suitable set of parameters for applying the Clean algorithm \citep{Hogbom1974,Schwab1984}, specifically determining two key values: how `deep' should we Clean (i.e., at what level of peak pixel value in the residuals map do we terminate the Clean process), and how the placing of model components should be constrained via the Clean mask. 
When performing a blind search for transients, unconstrained or `open' Clean reductions are preferable, since we do not know where a transient source may appear, and an `un-Cleaned' source will be harder to detect since the flux remains spread over the core and sidelobes of the dirty beam. 
As such, when imaging a field of view which contains no detectable steady sources (to search for transient sources), a simple approach to automating this process is to estimate the root-mean-square noise-level (RMS) from the dirty map, and then apply an open-Clean process, with the Clean termination threshold set to a multiple of the estimated RMS (typically around three).\footnote{%
Alternatively, if the performance of the telescope is well characterised, the visibilities may be analyzed directly to determine the total integration time excluding flagged data, and this value may be used to directly estimate the the RMS noise level. 
However, `well characterised' can be a difficult ideal to attain when telescope sensitivity may depend on miriad factors such as atmospheric moisture level, local ambient temperature, etc, and even then image-RMS may differ depending on proximity to a bright source.
}
However, this simple approach produces poor results when a bright steady source is located in the field of view, particularly if the single-epoch data being imaged is from a short integration with poor \textit{uv}-plane coverage.
In this case, the RMS as estimated from the dirty map will often be positively biased, due to additional pixel-value variance from the side-lobes of the bright source. 
Additionally, an open-Clean process may erroneously place model components in the side-lobes of the bright source, leading to artifical source-artefacts and reduced flux assigned to the model component at the true location of the bright source \citep[a phenomenon known as `clean bias', see e.g.][for details]{Condon1998,White1997}.
Therefore, when imaging a field containing a known bright source a better strategy is to perform a constrained (`masked') Clean --- but this potentially leaves transient sources `un-Cleaned', as noted previously. 

Previous radio-transient surveys have tackled this problem in different ways. \citet{Bell2014} performed phase and amplitude self-calibration on any standard calibration sources in the field of view, modelled and subtracted the contribution to the visibilities from those calibrators, and then finally applied an open-Clean process to the resulting image with a very limited number of Clean iterations, to avoid over-cleaning any remaining sources. 
This is effective at reducing side-lobes from calibration sources but will produce sub-optimal results if a bright source is present which is not identified as a calibrator. 
Alternatively, a manual round of source-identification may be employed prior to performing a final masked-Clean \citep{Miller2008,Mooley2013}, which is undoubtedly effective but can quickly become labour-intensive in the case of transient surveys, and is unsuitable for near-real-time analysis in any case.

The algorithm encoded in the \texttt{chimenea} package takes an iterative approach that makes use of all available epochs of data for a given survey field.
Initially, the data from all epochs are concatenated and an open-Clean process is applied, which is fairly robust to artefacts given enough epochs of data, due to improved \textit{uv}-plane coverage. 
To ensure the RMS-estimation (and hence Clean-termination threshold) is not biased by side-lobes in the dirty map, we re-estimate the RMS after applying an open Clean and then reapply the Clean process with a lower threshold if a significant drop in RMS is detected. 
The deep image synthesized from the concatenated data is then used to search for any steady sources in the field of view which are detectable with a high signal-to-noise ratio.
Next, each single-epoch observation is processed individually. 
We initially apply a masked Clean, to constrain the placement of Clean-model components to the location of any known bright sources in the field (as detected in the deep image) --- this avoids the issues of side-lobe artefacts due to poor \textit{uv}-coverage in shorter observations. 
Finally, we apply an open Clean to each epoch (initialized using the source-model components from the masked clean) in case a transient source is present at detectable flux-density levels.

\begin{figure}
\begin{center}  
  \includegraphics[width=0.45\textwidth]{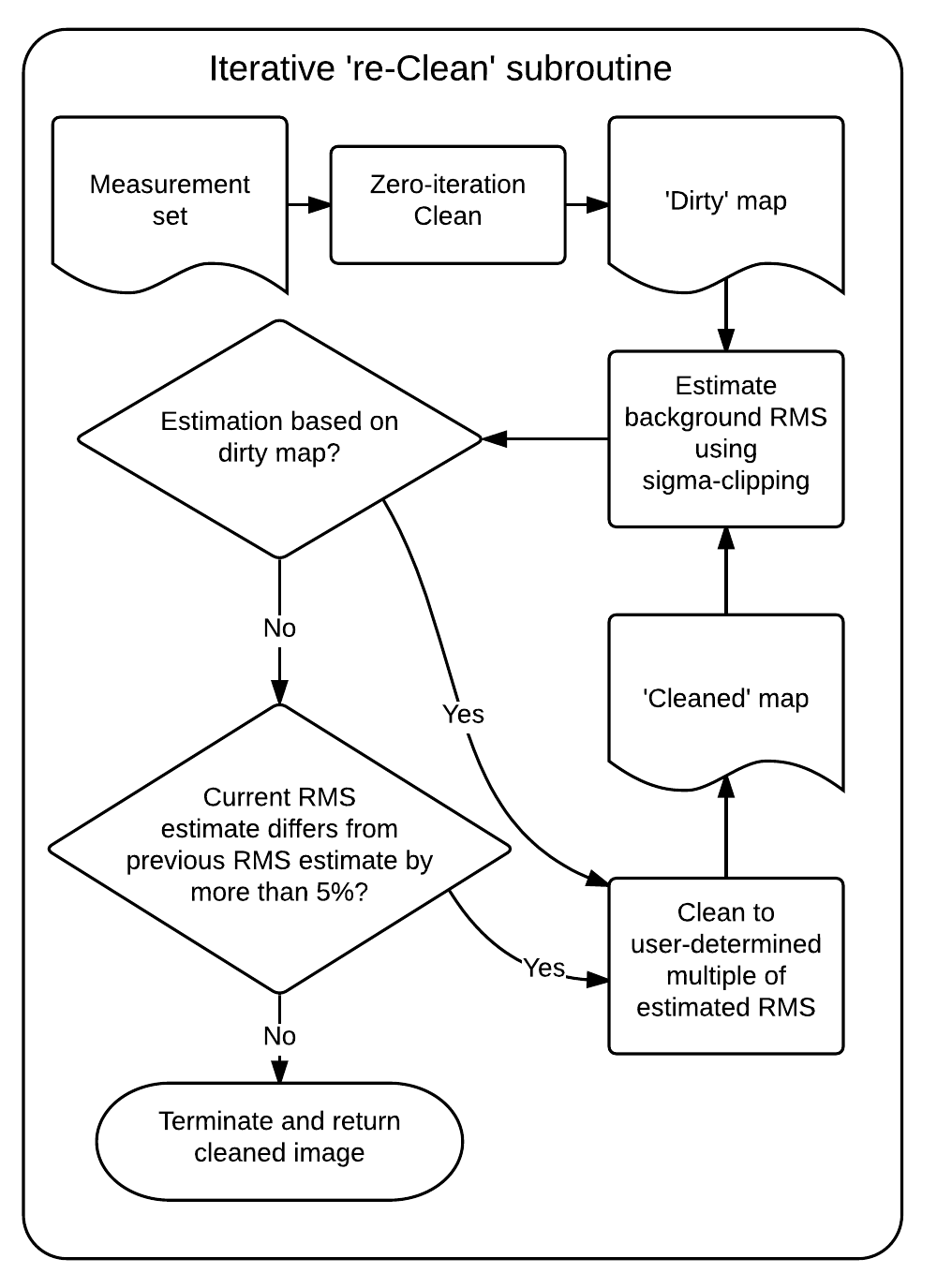}
  \caption[Iterative `re-Clean' subroutine logic.]{%
A schematic of the iterative `re-Clean' subroutine, applied as part of the \texttt{chimenea} algorithm depicted in Figure~\ref{fig:chimenea}.
The subroutine repeatedly applies the Clean algorithm until acceptance criteria are met. 
The convergence tolerance (shown as 5\% in figure), and the Clean threshold as a multiple of the estimated root-mean-square background noise level (RMS), are both user-configurable. 
An additional termination criterion is applied --- maximum number of iterations around the Clean / re-estimate RMS loop (typically three) --- but is omitted from the chart for brevity.

\label{fig:reclean}
} 
\end{center} 
\end{figure}

Figure~\ref{fig:chimenea} provides an overview of the implemented data-flow. 
A step-by-step description of the process, relating to the different stages denoted in Figure~\ref{fig:chimenea}, is as follows:

\begin{enumerate}[I]
 \item Concatenating data from all epochs gives better \textit{uv}-coverage and produces a `deep' map with lower noise, which is ideal for reliably locating steady sources in the field of view. To generate this deep map we apply the Cotton-Schwab variant of the Clean algorithm \citep{Schwab1984} to the concatenated data, allowing unconstrained placing of model components within a box-region corresponding to the central quarter of the synthesized map (referred to as an `open Clean' in Figure~\ref{fig:chimenea}). 
Sources outside this central region are attenuated by a primary beam response factor of less than 1 in 20,000 relative to the pointing centre, and so a wider open-Clean region is unnecessary unless this margin happens to contain a source with flux level on the order of one Jansky. 

 \item Applying the Clean algorithm requires that we give a pixel-value termination threshold, which is typically chosen as a multiple of the background root-mean-square (RMS) noise level. 
 However, in order to estimate the RMS, we require an image to sample. 
 To overcome this `chicken-and-egg' problem we take an iterative approach (`re-Clean'), as depicted in Figure~\ref{fig:reclean}. 
 Beginning with the dirty map, we take the pixel values from the residuals map (i.e. the image with any Clean-model components subtracted), apply sigma-clipping to estimate the background RMS,  apply Clean with the resulting calculated threshold, and then iterate by examining the new residuals map.
 Our chosen convergence criteria for ending the process is to stop when the proportional decrease in RMS estimates between iterations is less than a user-chosen value (typically 5\%) or when a set number of iterations have completed (typically three), whichever comes first. 
 For the sigma-clipping and RMS estimation step we employ the algorithm described in Section 4.3.1 of \cite{Swinbank2015}, which uses information on the synthesized beam shape to correct for bias in the RMS estimate caused by inter-pixel correlation.
 
 \item Applying an initially unconstrained iterative re-Clean process to each single-epoch observation can give poor results, especially if bright steady sources are present in the field of view, as poor \textit{uv}-coverage can lead to misidentification of side-lobes as separate sources,  and over-estimation of background RMS levels. 
 However, since the deep image is generated from multiple concatenated observations it typically has much better \textit{uv}-coverage and sensitivity. Therefore, we use the deep image to identify steady sources and generate a Clean mask. 
 This mask is used when imaging single-epoch data to constrain the Clean process to place model components only in the vicinity of known sources. 
 To identify sources in the deep image we make use of a source-extraction tool tailored to radio-data and currently distributed as part of the LOFAR transients pipeline \citep{Trap2014,Swinbank2015}.
 The user can also specify co-ordinates of known sources to create additional mask apertures, in order to ensure proper cleaning of any known faint sources or transient candidates. 
 
 \item  Applying the constrained (via use of the Clean aperture-mask) re-Clean algorithm to the single-epoch observations is generally quite robust and provides a good estimate of the background RMS at each epoch, in addition to accurately cleaning sources previously detected in the deep image.
 
 \item We can now apply an unconstrained Clean operation to each single-epoch observation, down to the flux-threshold determined by the previous masked re-Clean process, in case there are single-epoch transient sources present which do not correspond to a mask aperture. 
 We initialize this Clean operation with the source-component model derived from the masked Clean process, which should help to ensure that any bright sources in the field are cleaned correctly, without spurious assignment of source-components to bright side-lobe artefacts.
 
 \item Finally, the pipeline applies primary-beam corrections, and outputs images in both CASA \textit{MeasurementSet} and FITS formats.
\end{enumerate}

We note that the data-flow currently implemented in \texttt{chimenea} has been guided by our experience in reducing the ALARRM datasets, and may change in the future as we gain further experience.
The current pipeline is already a considerable improvement (see Section~\ref{sec:results}) over our initial automated solution, 
which was to simply perform an independent open Clean on each single-epoch observation \citep{Staley2013}, 
but particularly challenging or scientifically significant datasets will still warrant manual care and attention; \texttt{chimenea} provides a reduction which can sometimes be improved further but is good enough for scientific analysis in most cases.
Additionally, \texttt{chimenea} represents significant value as a working software-implementation, complete with extensive logging, a clearly structured logic-flow, and a number of simple data-structures which aim to make tractable the problem of managing numerous intermediate datasets and configuration parameters. 
We hope that others will be able to use and build upon this initial implementation.

To encourage re-use, the \texttt{chimenea} package has been kept as simple as possible, consisting solely of some basic data structures and the subroutines that make up the algorithm described above, with the complexities of calling out to \texttt{casapy} delegated to the \texttt{drive-casa} layer. 
However, this minimalist approach means that there is no command-line interface or default parameter set. 
These needs are met by the \texttt{AMIsurvey} package, described below.

%
%
\section{\texttt{AMIsurvey}, an end-to-end pipeline for automated reduction of AMI-LA observations}
\label{sec:amisurvey} 
\begin{figure}
\begin{center}  
  \includegraphics[width=0.4\textwidth]{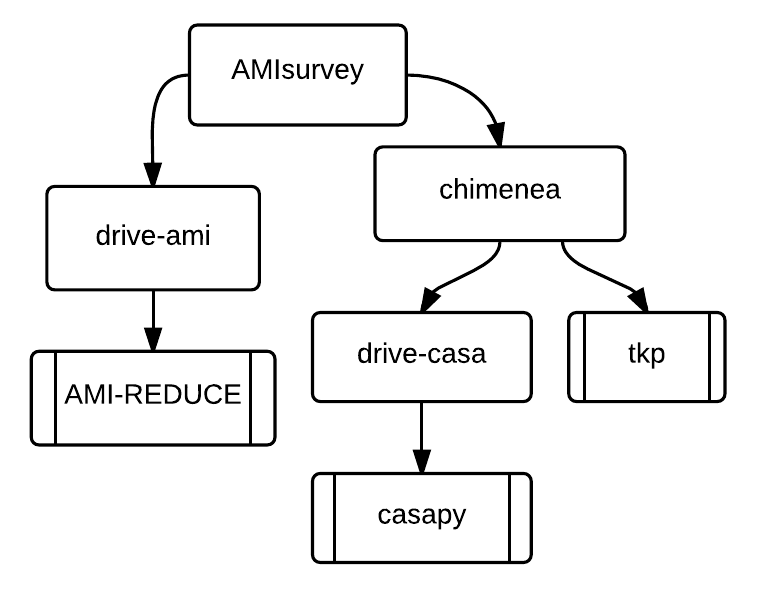}
  \caption[Dependency chart for the \texttt{AMIsurvey} package.]{%
A dependency chart for the \texttt{AMIsurvey} package; \texttt{AMIsurvey} depends on \texttt{drive-ami}  which in turn depends on the external package \texttt{AMI-REDUCE}, etc.
The \texttt{tkp} block refers to the LOFAR Transients Key Project software package \citep{Trap2014} described in \cite{Swinbank2015}.
Note that only the salient dependencies are listed; in reality we additionally make use of a 
number of external Python packages.
\label{fig:amisurvey-deps}
} 
\end{center} 
\end{figure}

The top-level tool used for data-reduction in the ALARRM programme is \texttt{AMIsurvey}. 
This package glues together the calibration and imaging processes implemented via \texttt{drive-ami} and \texttt{chimenea}, depending in turn upon the CASA and \texttt{AMI-REDUCE} packages (cf Figure~\ref{fig:amisurvey-deps}). 
We make use of the the Python standard library package \texttt{argparse} to provide a user-friendly command line interface.
Initially, the \texttt{AMIsurvey} tool loads lists of calibrated \textit{uv}-FITS files and associated metadata from the JSON files output by \texttt{drive-ami}. 
The data-locations and metadata are then trivially converted into the data-structure used by the \texttt{chimenea} pipeline, before being sorted into groups. 
These calibrated observations may be filtered according to quality-control criteria (e.g. limits on rain amplitude modulation) to identify and remove any problematic observations.
The grouped-and-filtered observations are then passed, along with configuration parameters (e.g. Clean image-synthesis settings) appropriate to the ALARRM data, to the main \texttt{chimenea} pipeline function for processing. 
Finally, a full listing of the data-products is output in JSON format, for easy ingestion by further analysis tools. 

\section{Performance testing and results}
\label{sec:results}
To thoroughly test the \texttt{AMIsurvey} pipeline, we reduced a significant portion of the ALARRM data obtained to date: 1035 observations spread over 165 target fields, representing approximately 100 days of telescope time with AMI-LA (approx. 100\,GB of raw data). 
The calibration and imaging stages took around eight and six hours respectively on a single CPU-core. 
The metadata associated with these reductions were then analysed to confirm that the pipeline generally behaves as expected, identify any problematic datasets, and place some quantitative values on the benefits of applying the full \texttt{chimenea} algorithm, as opposed to simply applying a single Clean to each observation.\footnote{%
 The \texttt{IPython} notebook used to perform this analysis is included with the \texttt{AMIsurvey} package, and a static version relating to the dataset described herein may be viewed at  \url{http://nbviewer.ipython.org/gist/timstaley/5ef6dfc2a370e0288bb0}.
}

The first step of the analysis was to exclude any fields of view for which no sources were detected in the concatenated deep-image at the 5.5 sigma level --- these fields have only a single open Clean operation performed at each epoch, based on the RMS estimated in the dirty map, and so do not provide relevant datapoints for this comparison.
This left 862 observations spread over 121 target fields where a source was detected in the deep image.
The data were reduced with (somewhat arbitrarily chosen) re-Clean convergence criteria of a RMS-decrease less than 5\% after applying a Clean operation, or a maximum of three re-Clean iterations (cf Figure~\ref{fig:reclean}). 
For the ALARRM dataset, these values appear quite suitable --- 503 observations converged after one iteration (i.e. less than 5\% difference in estimated RMS between dirty map and image produced from one masked Clean operation), 296 converged after two iterations, leaving 63 observations for which the re-Clean algorithm iterated the maximum of three times. 
Of these 63 remaining observations, 62 had an RMS decrease in the final re-Clean iteration of less than 5\%. Therefore we can consider only one observation as requiring a `hard-stop' on the re-Clean iterations. 
(This non-converging observation turns out to be from a field with extended emission, which is generally problematic for the current \texttt{AMIsurvey} pipeline.)

\begin{figure*}[p]
\begin{center}  
  \includegraphics[width=0.8\textwidth]{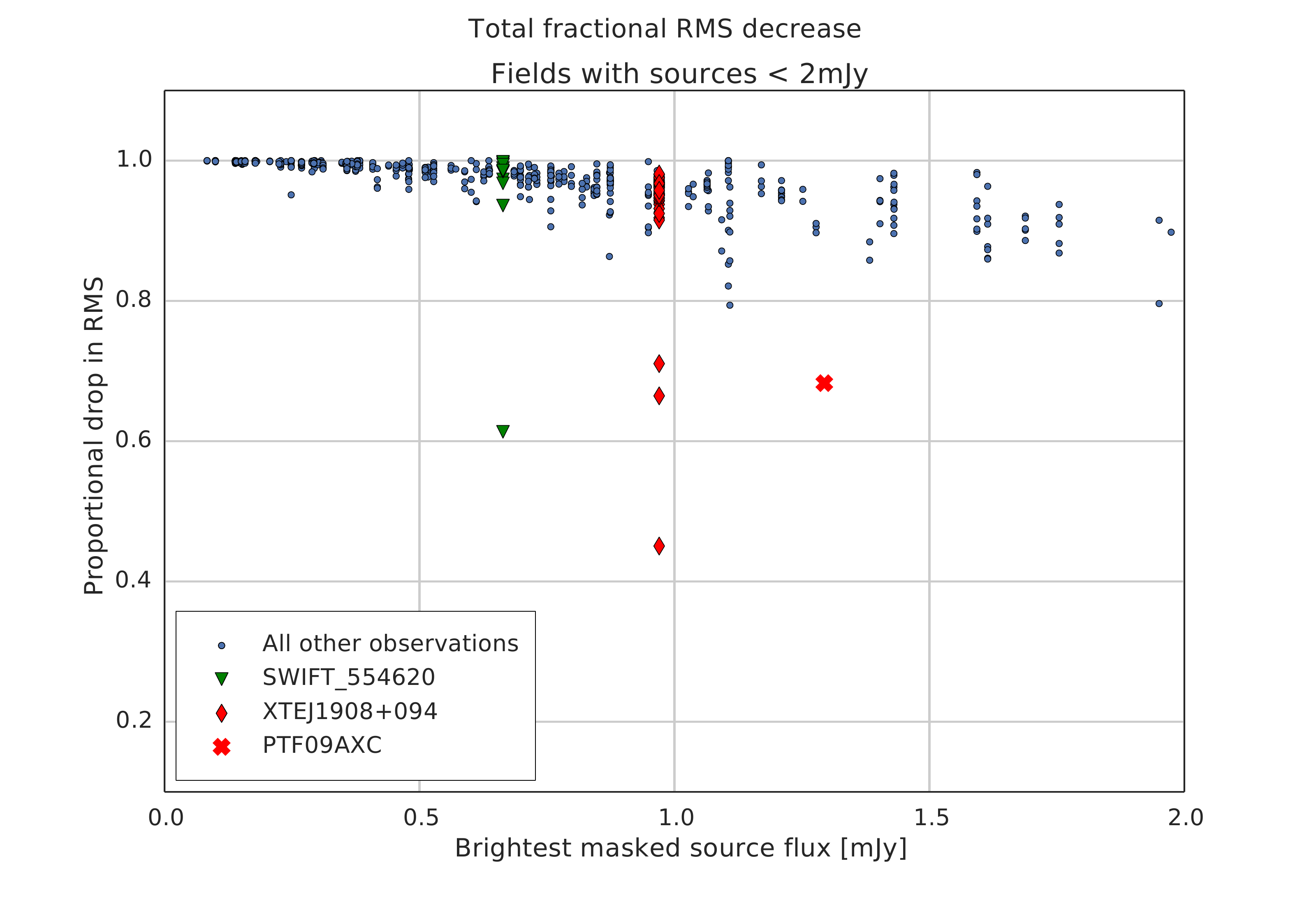}
  \includegraphics[width=0.8\textwidth]{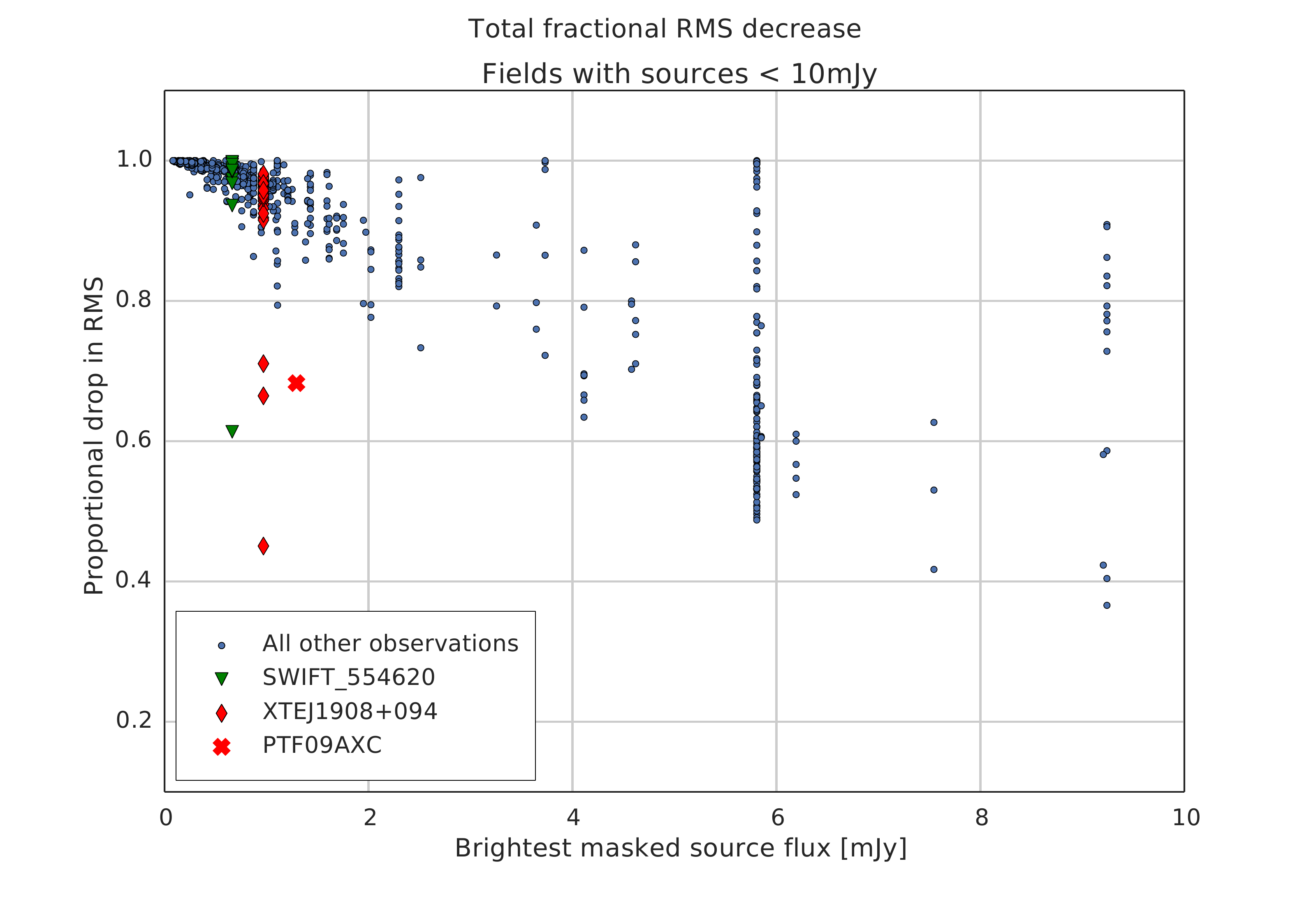}
  \caption[Fractional RMS decrease vs. brightest source flux]{%
  \label{fig:rms}
  Scatter-plots comparing the fractional decrease in RMS-estimate (the ratio of [final image RMS / dirty map RMS] for each single-epoch observation), and the flux-density from the brightest source-detection in the deep image of that observation's field of view. 
  Each target has a single value for `brightest source-flux in deep image,' but RMS-decrease varies between epochs, resulting in all RMS-decrease values for a single target being plotted as a column of points.
  As expected, fields with brighter sources in view often have more drastically reduced RMS levels than fields with only faint sources.
  A few outliers (overplotted in red and green) buck this trend, for well understood reasons --- details in text.
  Upper and lower sub-plots depict the same data but with different limits on plotting region applied, to give more clarity in the densely populated upper-left region.
} 
\end{center} 
\end{figure*}

\begin{figure*}[p]
\begin{center}  
  \includegraphics[width=0.8\textwidth]{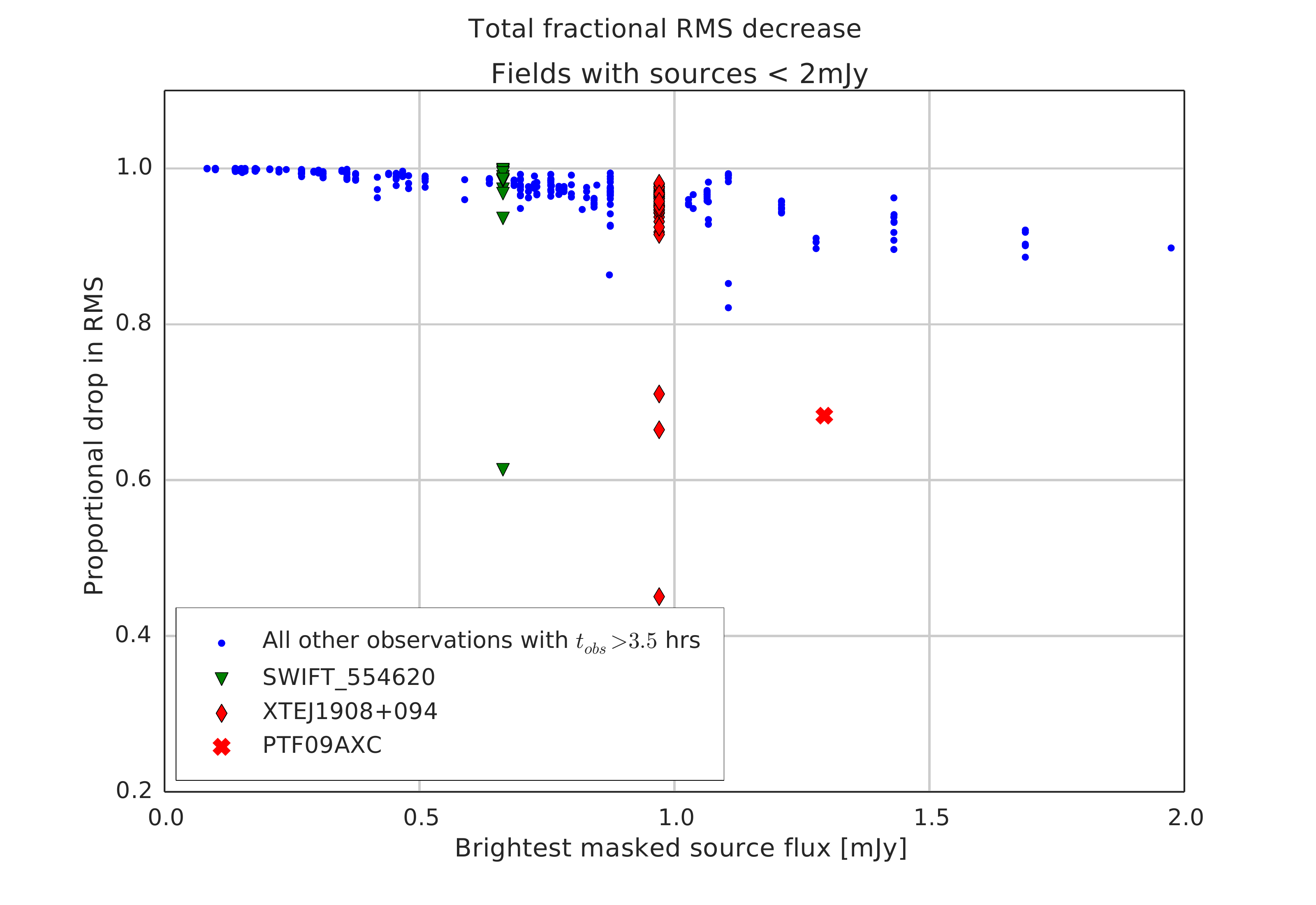}
  \includegraphics[width=0.8\textwidth]{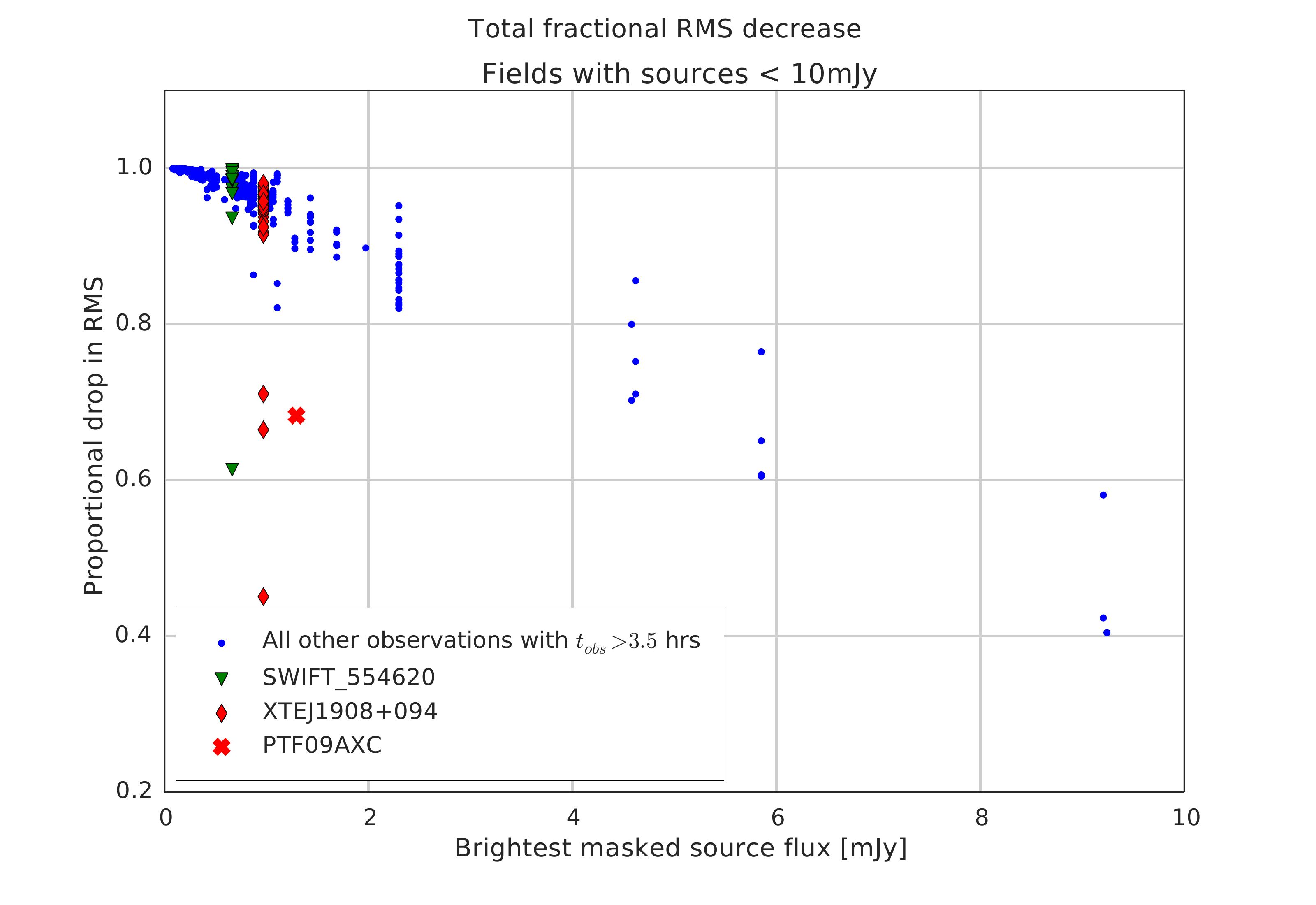}
  \caption[Fractional RMS decrease vs. brightest source flux for long-duration observations]{%
  \label{fig:rms_long_duration}
  As Figure~\ref{fig:rms}, but this time only observations of duration 3.5 hours or greater (typically 4 hours) are plotted. 
  With the short-duration observations removed, the plots show a clearer relation between the flux-density level of the brightest source in the field, and the RMS-estimation improvement gained by applying an iterative Clean process.
} 
\end{center} 
\end{figure*}

As an additional check on the pipeline's behaviour, we produced Figures~\ref{fig:rms}~and~\ref{fig:rms_long_duration}, which compare the fractional decrease in RMS-estimate between the dirty map and the final image, and the flux-density of the brightest source detected in the deep image. 
Generally, we expect fields with brighter sources to exhibit higher levels of side-lobe artefacts in the dirty map, which positively bias the initial dirty-map RMS-estimate (particularly for short observations with poor \textit{uv}-coverage), and so we expect some level of correlation between these two quantities. 
The resulting plots show a smooth progression in the maximum RMS-decrease with increasing source flux, excepting some well-understood outliers:
\begin{itemize}
   \item The observations of XTEJ908+094 are affected by a very bright out-of-field source, which presents extended side-lobe artefacts as additional background variation in the field-of-view.
   \item The PTF09AXC field contains extended emission and/or blended sources (which are not well suited to reduction by the automated pipeline).
   \item The observations of field SWIFT\_554620 contain a single epoch with a particularly bright source (GRB140327A in afterglow), and so this observation falls lower on the  plot than the rest of the grouping.
\end{itemize}
In total there are 84 observations of fields with sources brighter than 10\,mJy, which are not shown in Figures~\ref{fig:rms}~and~\ref{fig:rms_long_duration}.
When plotted, these datapoints continue the general trend with increased scatter, but we do not consider them as suitable for automated reduction in the current \texttt{AMIsurvey} pipeline as modified calibration settings need to be applied when a source of around 10\,mJy or brighter is in the field of view.

As can be seen in Figure~\ref{fig:rms}, when a bright source is present the RMS-estimates may drop by a factor of two or more after the masked re-Clean process is applied (i.e. a proportional drop to less than 0.5 times the original value, in the units plotted). 
Since this RMS-estimate is then used when applying the open-Clean process to look for single-epoch transients, it suggests a potentially significant improvement in sensitivity to transient sources when a bright steady source is present in the field of view. However, some observations of fields with a bright steady source do not show such improvement. This is to be expected if short observations, or those which undergo a high degree of RFI flagging, have sufficiently poor data quality that bright source sidelobes are not the dominant contributing factor to background variation. To test this hypothesis, we made a simple cut on the data by replotting only observations of duration greater than 3.5 hours, as shown in Figure~\ref{fig:rms_long_duration} (leaving 289 observations spread over 65 different targets). This plot shows a much clearer relation between bright-source flux-density and RMS-estimate reduction.

We note that the RMS reduction factors (between dirty-map RMS-estimate and final RMS-estimate) displayed in Figures~\ref{fig:rms}~and~\ref{fig:rms_long_duration} give an indication of the improved imaging quality of the \texttt{AMIsurvey} pipeline compared to our early reductions, since the naively-reduced images would be cleaned to a threshold determined by the dirty-map estimated RMS. 
The addition of a masked-Clean process and output image should also help to avoid under-estimation of flux in bright sources (`Clean-bias').
However, the ultimate figures-of-merit relevant to transient surveys are those relating to sensitivity and accuracy in transient detection. 
Experience suggests the trade-off between sensitivity to real transients and artefacts in radio-astronomy is difficult to quantify without extensive testing against both real and simulated visbility data, which we leave to future investigation.

\section{Discussion}
\label{sec:discussion}
\subsection{Limitations, possible extensions and alternatives to \texttt{chimenea}}
\label{sec:chimenea-discussion}
While \texttt{chimenea} and the other packages presented here have proven useful in the reduction of ALARRM datasets, the approach is not without its limitations.
The most immediate problem is that the RMS-estimation procedures do not work well in the presence of extended emission or complex blended sources, and so fields containing such sources still need to be manually reduced (though these types of source are always tricky when reducing radio-synthesis data). 
The iterative approach also adds a non-trivial amount of computational overhead, since a field of view may be processed with a Clean operation multiple times as the background RMS is re-estimated. 
The ALARRM datasets to date have been relatively quick to reduce, with file sizes on the order of a gigabyte or less and reduction times on the order of minutes, but for a large dataset the re-Cleaning may become a significant contribution to the overall processing time. 

We should also reiterate that this work has been motivated to date by reproducing the data-reduction steps we would previously have applied manually. 
With this benchmark implementation in place, we can begin to consider alterations which might improve sensitivity and accuracy. 
A logical extension would be to utilise \textit{uv}-plane model fitting, to improve the accuracy of flux-density measurements after candidate sources have been identified via the current pipeline. 
This avoids the degeneracies involved in `fitting a model to a model' --- i.e. fitting a source-model to a synthesized image that has already been produced by inference from the raw data  \citep{MartiVidal2014}.
Additionally, it would be interesting to compare the results from our pipeline to some of the more recent source-detection and modelling approaches that have been developed on the basis of Bayesian inference \citep[e.g.][]{Sutter2014,Lochner2015}, since these provide a full modelling of the data from the \textit{uv}-plane information, although a significant trade-off in terms of increased computational overhead is to be expected. 
Image-plane transient-detection pipelines for the SKA will have stringent demands on both accuracy and computational budget, and so it may be that some hybrid or simplified image-synthesis approach will turn out to be most attractive in the long run.

\subsection{Methods for interfacing with legacy or proprietary software}
\label{sec:software}
The \texttt{drive-ami} and \texttt{drive-casa} software packages discussed in this paper solve a problem commonly found throughout software-intensive fields of both academia and industry --- how to employ (in an automated fashion) a software package that cannot be directly called as a library. 
This may arise when the external package is written in a different programming language to that of the user's codebase, or because the external codebase is legacy or proprietary and only the command-line interface is accessible.
In this subsection we describe three approaches that may be adopted when creating an interface to legacy or cross-language software, together with their merits and drawbacks, and explain why we chose to use terminal-emulation in this work.

\subsubsection{Cross-language bindings}
Typically, the core algorithms of data-reduction pipelines are 
implemented in a lower-level language such as Fortran, C or C++, either for 
performance reasons or simply because these languages were most suitable
when the software was initially written. However, it is generally recognised
that high-level programming languages such as Python, Perl or IDL provide a more
suitable environment for rapid prototyping and development of scientific 
algorithms, due to their flexibility, succinctness, and 
convenient access to a large variety of external libraries for tasks such
as data-visualization or statistical analysis.
The standard approach to making routines written in a low-level language 
available from a higher-level language is to 
implement cross-language ``bindings'', 
specialized modules which map data-structures and function calls between the two.
Examples of this approach in radio 
astronomy are \texttt{miriad-python} \citep{Williams2012}, which provides 
Python bindings for \texttt{miriad} \citep{Sault1995,Sault2011},
and the \texttt{Obit / ParselTongue} packages 
\citep{Cotton2008,Cotton2013,Kettenis2006, Kettenis2012}, which provide 
Python bindings for AIPS \citep{Fomalont1981}. 
We refer the reader to \cite{Williams2012} for further discussion on these
packages and their merits.

Cross-language binding modules are usually the most desirable solution for developing a scriptable interface to routines written in a low-level language, since a well-written set of modules will provide a high-performance, seamless interface between the external package and the user's programming language of choice.
However, there are associated costs and requirements that may make such an approach impractical. 
In astronomy, the primary issues with software development are usually those of available developer time and expertise.
Astronomers with knowledge of both high-level and low-level languages, and how to implement bindings between them, are rare. 
Even if someone with the requisite skills is available, the task of creating bindings and porting the required pipeline-logic to a high-level language may require a significant investment of developer time, both to understand the logic implemented in the legacy system, and to verify that it is faithfully reproduced in the `ported' version. 
At the end-user level, there is also the problem of familiarity: a cross-language port of a well-used legacy package will require users to re-learn the interface via its new bindings. 
Finally, there are the twin issues of access to source code, and maintenance --- 
if a tool is proprietary, or if the source is unavailable for other reasons, or non-trivial to compile, then this presents an additional hurdle to implementation and use of bindings. 
Any changes to the original package must also be reflected in the bindings modules, requiring code-maintenance and recompilation on a schedule tied to the external package.

As such, implementing binding modules from scratch may be 
overkill.
Some alternative approaches to using low-level or legacy codes from a higher-level language are described below.

\subsubsection{Sub-process calls}
\label{sec:subprocess}
If an external tool is usually invoked via an executable which takes command line arguments and then requires no further interaction, it is often trivial to invoke in a scripted fashion (otherwise known as `spawning' or `calling out' to a sub-process). 
The built in Python library \texttt{subprocess} provides a perfectly adequate interface when this is the case. 
We also direct the reader towards the \texttt{sh} 
package\footnote{
\url{http://amoffat.github.io/sh/}
}, which 
provides an alternative interface in which external commands may be represented by objects akin to Python functions, making for a neater syntax in repeated usage.

However, this approach falls down when invoking longer-running, 
multi-step processes that may require user interaction. 
While such tools often provide an option
to run a series of commands from a script, parsing the concatenated output 
from many individual commands to record details of the data-reduction, 
or worse yet detecting error conditions and locating the cause, 
can present a considerable challenge. 
In addition, larger, interactive tools often incur a few seconds start-up
time while the environment loads, and so repeatedly invoking the tool for 
multiple smaller steps can be impractically slow. 

\subsubsection{Terminal emulation}
\label{sec:emulation}
A third option is that of terminal emulation: spawn the external tool in a sub-process, but then interact with it by emulating interactive terminal input.
While this is slightly more involved than a simple sub-process call it allows for a one-to-one mapping of the sub-routines presented by the external tool into the user's scripting language of choice, without any of the source modification or recompilation that implementing bindings requires. 

In Python, terminal emulation functionality can be provided by a package called \texttt{pexpect}\footnote{%
\url{https://github.com/pexpect/pexpect}%
}.
The \texttt{pexpect} package file is small (a 132kB download) and only requires a working Python installation (v2.6 or greater) as a prerequisite. 

When building an interface to a legacy tool using \texttt{pexpect} in Python, we have found that it usually makes sense to provide an interface class. 
This class can be used to spawn the external process when initialized, store a reference to the \texttt{pexpect} object representing that process and the emulated terminal interface, handle the terminal interaction, and so on. 
Implementation of the basic interface class is relatively simple. 
The essential parameters are a regular expression representing the format of the command-line prompt presented by the external tool, the path to invoke the external tool, and any required environment variables.
Once the underlying interface is in place, end-user scripts that would previously be invoked from within the external tool's interactive environment can immediately be employed as part of a larger pipeline. 
Furthermore, the precise inputs passed via the interface can be recorded alongside any data-outputs, providing data-provenance, so that if an end-user later wishes to inspect the data-reduction process manually, they can load up the external tool and step through the recorded command-script, either verifying or varying the reduction steps as required.

From this starting point, it is straightforward to build up a more comprehensive interface suitable for use in complex logic flows. Since the output from each command is available in a clearly separated block of text, it is often trivial to implement basic parsing routines that can recover information from the command-line output of the external tool, allowing for `if / else' logic branches in the larger pipeline which make use of intermediate outputs from the external tool. 
If desirable, convenience routines can be added to ease the process of generating the commands passed to the external tool, providing an interface which is better suited to the high-level scripting language (while still allowing direct access to the external tool as required).

Since this approach effectively treats an external tool's command-line interface as a programming interface, any updates to the external tool are unlikely to require changes to the scripting-interface, unless they involve changes that also change the user-interaction behaviour (with one caveat --- if output-parsing relies on a particular output format, this may be changed to another layout which is still human-friendly but breaks the fixed parser rules).

In terms of performance, terminal emulation sits somewhere between that of cross-language bindings and simple sub-process invocation. 
Small delays (typically around 0.05 seconds) in the input-output cycle must be allowed for when emulating terminal interaction, presumably due to the small but measurable time it takes for the external tool to parse the commands input. In addition, any overhead incurred when starting the external tool is still present. 
However, once initialized, an instance of the external tool can be re-used for many commands, and so this start-up overhead is incurred fewer times when compared to simple sub-process calls.

It is this terminal-emulation approach that we adopted in implementing Python interfaces to the \texttt{AMI-REDUCE} and CASA packages, for reasons outlined below.

\subsubsection{Use of terminal-emulation in \texttt{drive-ami} and \\ \texttt{drive-casa}}
\label{sec:why-emulation}
The \texttt{AMI-REDUCE} pipeline is implemented in \mbox{FORTRAN} 77, and comprises around 58,000 lines of 
source-code\footnote{As analyzed using the \texttt{SLOCCount} utility: \\ \url{http://www.dwheeler.com/sloccount/}\\
(Including component libraries.)}. 
In the rare case that \texttt{AMI-REDUCE} needs to be run elsewhere than the MRAO it is typically distributed as a pre-built binary, with a copy of the source files included. 
Despite the author's efforts to install appropriate dependencies and alter makefiles, building the package from source on a modern Ubuntu Linux installation proved to be problematic (as opposed to simply executing the pre-built binary, which was simple once the appropriate compatibility packages had been downloaded and installed). 
Given the size and complexity of the codebase, the difficulties building from source, and the authors' lack of expertise in Fortran, we judged that implementing Python-bindings to the \texttt{AMI-REDUCE} \mbox{FORTRAN} routines directly would require a significant and costly investment of developer time, with little benefit given that \texttt{AMI-REDUCE} is not a widely-used software package. 
As described in Section~\ref{sec:drive-ami}, simple sub-process calls to \texttt{AMI-REDUCE} were initially used in our end-to-end pipeline, but soon proved inadequate. 
Terminal-emulation presented a compromise - greater control and flexibility in interfacing with \texttt{AMI-REDUCE}, without requiring a large effort to create \mbox{FORTRAN}/Python binding modules. 
The added logging capabilities also provide reassurance that our data-reduction process is performing as expected, and can be verified manually if required.
Development of the \texttt{drive-ami} interface using terminal emulation allowed for a relatively low initial time-investment to create a basic interface class. 
Additional features such as error-handling and metadata-extraction were then added piecemeal on an `as needed' basis, since addition of minor features can typically be accomplished in a few hours.

The case for writing a terminal-emulation tool for CASA is less obvious, since \texttt{casapy} already provides a scriptable Python interface --- on the face of it, writing a Python library to interface with a Python interpreter is faintly ridiculous.
However, as detailed in in Section~\ref{sec:drive-casa}, use of CASA routines alongside other Python libraries is otherwise a difficult and error-prone process. 
In the long-term this may change, but for now the additional logging capabilities, together with the fact that \texttt{drive-casa} can be trivially installed on most systems, mean that it provides significant benefits to the end-user in terms of both convenience and reproducibility.

\subsection{Writing re-usable Python codes}
\label{sec:code-reuse}
This work highlights a pressing issue for the field of astronomy as a whole (and wider academia): that of software interoperability and re-use. 
There will always be a trade-off to be made in choosing to incorporate legacy packages into new efforts, or starting afresh with a re-write in the language \textit{du jour} --- nothing has changed there. 
However, when writing new software, we would encourage authors to consider how their code might be adapted for re-use, either in part or in whole. 
In the wider context of software development, situations are so varied that it is hard to give any firm advice on this topic (or at least,  beyond the scope of this work). 
However, if we restrict our scope to that of writing scientific codes in Python, we can be a little more concrete; giving a few examples from this work that hopefully serve to illustrate the wider argument:
\begin{itemize}
 \item Prefer interoperable, modular libraries over monolithic pipeline packages 
  (this is a widely discussed viewpoint more generally known as the \textit{Unix philosophy}). 
  The packages described in this paper began life as one big collection of prototype code that called out to each external tool in turn (via a sub-process call).
  As we added features and the codebase grew, it became clear that it was worth separating into separate packages.
  
 \item Use the Python packaging infrastructure. 
  A major hurdle to breaking up a code into sub-packages is that it induces additional dependency management.
  Embracing the Python packaging infrastructure to implement a versioned install process helps keep this under control, and also makes it much easier to distribute the code to users not directly involved in the development 
  process.\footnote{A guide to creating installable Python packages may be found at \href{http://the-hitchhikers-guide-to-packaging.readthedocs.org/}{\url{http://the-hitchhikers-guide-to-packaging.readthedocs.org/}}}

 \item Ideally, distinguish technical implementation details from higher-level applications of code (or `low-level' vs `high-level' programming interfaces). 
    Core libraries should ideally be simple (each function does one thing well) and flexible (functions can be used independently, perhaps requiring a number of parameters, etc).
  However, when writing a larger software package it is essential to hide common-usage patterns of a low-level interface behind larger subroutines - this is the only way to write complex code which is readable and maintainable. 
   Think carefully about where it might be sensible to divide these code hierarchies and separate them into high-level and low-level packages. For example, in this work \texttt{drive-casa} provides a direct interface to the \texttt{casapy} package (low-level), while \texttt{chimenea} encodes the scientifically relevant logic and implements a particular view of how best to structure the data (high-level).

 \item Avoid use of the \textit{print} function. 
    The Python \texttt{logging} package provides a good degree of control and flexibility in how debugging and information messages are displayed, stored to file, distributed over the network, etc, whereas the \textit{print} function is a basic on/off output and should only ever be used as a temporary measure --- otherwise you risk overwhelming calling routines with a flood of irrelevant debugging output.
    There are also many extensions to the \texttt{logging} functionality, for example \texttt{AMIsurvey} makes use of the 
  \texttt{colorlog} package\footnote{\href{https://github.com/borntyping/python-colorlog}{\url{https://github.com/borntyping/python-colorlog}}} 
to colourise the terminal output by logging category.

 \item Separate configuration from code.
   Ideally, any scientifically significant parameters should be clearly identifiable and easily changed from one location. 
   In the current \texttt{AMIsurvey} pipeline, telescope and data specific parameters are encoded in a single, short module dedicated to the AMI-LA telescope. 
   This could easily be extended to load a different module depending on metadata entries, or load parameters from a user-edited text file as required.
\end{itemize}

Finally, we highly recommended that even if a scientific code follows none of this advice, it should still be made freely available. 
Stylistic changes to suit a given context can always be made after release, and the benefits of source code sharing are significant, perhaps most crucially for encouraging reproducibility \citep[see e.g.][for further discussion]{Shamir2013}.

\section{Conclusions}
\label{sec:conclusion}
In this work we have described an automated pipeline for reduction of radio-synthesis data obtained for transient follow-up, and undertaken inital performance tests which show it behaves as desired. 
Although heuristic in approach, \texttt{chimenea} already provides a useful component in automatic data-reduction pipelines, and serves as a test-bed and benchmark for more sophisticated reduction algorithms. 
We have highlighted limitations to the pipeline and discussed possible avenues for future development and exploration.
We note that extensive further development in the area of automated imaging will be essential for performing real-time image-plane transient surveys, one of the key science drivers for the SKA \citep{Burlon2015,Corbel2015,PerezTorres2014}.

From a technical standpoint, we have demonstrated the effectiveness of terminal-emulation as a method for interfacing with legacy or externally maintained software.
We found this approach relatively easy to implement and well suited to our needs, especially given the added benefits of easy manual reproducibility and data-provenance.  

Finally, we have provided some basic guidelines and examples towards writing re-usable Python codes in astronomy. 
This topic could be the subject of a much larger work, but we hope the suggestions herein will stimulate further thought and discussion. 
In particular, we hope that our efforts to separate and document the components of the \texttt{AMIsurvey} pipeline will result in productive re-use elsewhere.

%
%

\section*{Acknowledgements}
We thank the staff of the Mullard Radio Astronomy Observatory for their invaluable assistance in the operation of AMI.

This work made use of a number of publicly available software packages and data products: NASA's Astrophysics Data System; Astropy, a community-developed core Python package for Astronomy \citep{Astropy2013}; the IPython package \citep{PER-GRA:2007}; matplotlib, a Python library for publication quality graphics \citep{Hunter2007}; and SciPy \citep{jones_scipy_2001}. 

This project was wholly funded by European Research Council Advanced Grant 267697 
`4 $\pi$ sky: Extreme Astrophysics with Revolutionary Radio Telescopes.’

\section*{References}
\bibliographystyle{elsarticle-harv-max8}
\bibliography{automated_radio_reduction}

\end{document}